\documentclass[12pt,preprint]{aastex}

\newcommand{\kev}{keV}
\newcommand{\fe}{Fe~K$\alpha$}

\newcommand{\mcg}{MCG--6-30-15}

\shorttitle{The Cosmic Density of Relativistic \fe\ Lines}
\shortauthors{Ballantyne}

\begin{document} 

\title{The Distribution and Cosmic Density of Relativistic Iron Lines
  in Active Galactic Nuclei}


\author{D. R. Ballantyne}
\affil{Center for Relativistic Astrophysics, School of Physics,
  Georgia Institute of Technology, Atlanta, GA 30332}
\email{david.ballantyne@physics.gatech.edu}

\begin{abstract}
X-ray observations of several active galactic nuclei show
prominent iron K-shell fluorescence lines that are sculpted due to
special and general relativistic effects. These observations are
important because they probe the space-time geometry close to distant
black holes. However, the intrinsic distribution of Fe line strengths
in the cosmos has never been determined. This uncertainty has
contributed to the controversy surrounding the relativistic
interpretation of the emission feature. Now, by making use of the
latest multi-wavelength data, we show theoretical predictions of the
cosmic density of relativistic Fe lines as a function of their
equivalent width and line flux. We are able to show unequivocally that
the most common relativistic iron lines in the universe will be
produced by neutral iron fluorescence in Seyfert galaxies and have
equivalent widths $< 100$~eV. Thus, the handful of
very intense lines that have been discovered are just the bright end
of a distribution of line strengths. In addition to validating the
current observations, the predicted distributions can be used for
planning future surveys of relativistic Fe lines. Finally, the
predicted sky density of equivalent widths indicate that the X-ray
source in AGNs can not, on average, lie on the axis of the black hole.
\end{abstract}


\section{Introduction}
\label{sect:intro}
The X-ray spectrum of many active galactic nuclei (AGNs) exhibits a strong emission feature at
an energy of 6.4~\kev\ that is due to the fluorescence of weakly
ionized iron in a dense and relatively cold medium
\citep[e.g.,][]{nan89,pou90,np94}. The strength of the emission feature
requires that the region responsible for the fluorescence subtend
about half the sky as seen from the illuminating X-ray source
\citep[e.g.,][]{gf91}. As the variability properties of AGN place the X-ray
source within 10~Schwarzschild radii from the black hole \citep[e.g.,][]{gra92,utt07},
it is likely that relativistic effects may alter the observed line
shape \citep{fab89,laor91} allowing it to be a powerful probe of
space-time curvature. Several examples of
relativistically broadened \fe\ lines have been detected in the
spectra of AGNs over the last decade \citep[e.g.,][]{tan95,fab02,rn03,jmm07,fab09}. These
detections are leading to measurements of black hole spins and
accretion disk dynamics \citep{br06,rf08}.

The strength of the Fe line emission depends on only four parameters:
(i) the shape of the illuminating continuum \citep{bfr02,br02}, which in
AGNs is a power-law with photon index $\Gamma$; (ii) the abundance of
iron (relative to the solar value) in the accretion disk \citep{bfr02},
$A_{\mathrm{Fe}}$; (iii) the ionization parameter of the illuminated
disk \citep{bfr02}, $\xi=4\pi F_{X}/n_{\mathrm{H}}$, where $F_{X}$ is
the illuminating flux, and $n_{\mathrm{H}}$ is the density of the
illuminated slab; and (iv) the reflection fraction $R$, a measure of
the relative strength of the reflection component in the observed
spectrum (this is related to the covering factor of the accretion
disk). Recently, multi-wavelength survey data has been able to measure
correlations between each of the first three of these parameters to
$\lambda=L_{\mathrm{bol}}/L_{\mathrm{Edd}}$, the Eddington ratio of
AGNs \citep{rye09,ith07,nt07}. Here, $L_{\mathrm{bol}}$ is the
bolometric luminosity of an AGN, and $L_{\mathrm{Edd}} = 1.3\times
10^{38}\ (M_{\mathrm{BH}}/M_{\odot})$~erg~s$^{-1}$ is the Eddington
luminosity of a black hole with mass $M_{\mathrm{BH}}$ and $M_{\odot}$
is the mass of the Sun. In addition, various surveys have been able to
measure the black hole mass distribution \citep{net09} and its evolution
with redshift \citep{lab09}, as well as the luminosity function of AGNs
as a function of redshift \citep[e.g.,][]{ueda03}. We can then combine all this
information to calculate the density and evolution of the \fe\ line
from accretion disks. In the following we concentrate solely on the
\fe\ line that arises from the inner accretion disk, and neglect the
contribution from any narrow component \citep[e.g.,][]{nan06}, or one that arises
from reflection off a Compton-thick absorber \citep[e.g.][]{my09}. The following
$\Lambda$-dominated cosmology is assumed \citep{sper03}:
$H_0=70$~km~s$^{-1}$~Mpc$^{-1}$, $\Omega_{\Lambda}=0.7$, and
$\Omega_{m}=0.3$.

\section{Calculations}
\label{sect:calc}
The 2--10~keV X-ray luminosity function measured by \citet{ueda03} is used as the
basis for the density and evolution of AGNs. We consider 261
X-ray luminosities between $10^{41.5}$ and $10^{48}$~erg~s$^{-1}$, and
assume a bolometric correction of $50$ to convert these to
$L_{\mathrm{bol}}$ \citep{vf07}. The black hole mass distribution at $z\approx 0.15$ measured
by \citet{net09} is converted into 28 mass bins (spanning
$M_{\mathrm{BH}}=1.7\times 10^6$~$M_{\odot}$ to $9.1\times
10^8$~$M_{\odot}$) and normalized to sum
to unity. Applying this distribution to each of the 261 luminosities
yields 261 Eddington ratio distributions. At $z=0.15$ the minimum and maximum
Eddington ratios are $\lambda=1.4\times 10^{-4}$ (at a X-ray
luminosity of $10^{41.5}$~erg~s$^{-1}$) and $2\times 10^5$ (at a X-ray
luminosity of $10^{48}$~erg~s$^{-1}$), respectively. A constant density ionized
reflection code is used to calculate the reflection continuum
including the \fe\ line \citep{rf93,rfy99,rf05}. In these models, the density of
the accretion disk is taken to be $n_{\mathrm{H}}=10^{15}$~cm$^{-3}$,
and $F_{X}$ is changed in order to obtain the desired value of
$\xi$. The observed relationships between $\xi$, $\Gamma$,
$A_{\mathrm{Fe}}$ and $\lambda$ \citep{rye09,ith07,nt07} are used to determine the input parameters for the
the reflection code over a wide range of $\lambda$. The parameters
are frozen at reasonable values\footnote{The upper limit of
  $\Gamma=2.55$ was chosen for computational reasons, but, despite the
  rare example \citep[e.g.,][]{fab09}, the vast majority of AGNs have
  photon indices less than this value \citep[e.g.,][]{rye09}.} for very high or low values of
$\lambda$. The relationships are:  
\begin{equation}
\Gamma = \left\{ \begin{array}{lcl}
                0.58(\log \lambda + 1.0) + 1.99 & \phantom{xxxxxx} &-2.0 \leq \log
		\lambda \leq -0.035 \\
		2.55 & \phantom{xxxxxx} & \log \lambda > -0.035
		\end{array}
                \right.
\end{equation}
\begin{equation}  
A_{\mathrm{Fe}} = \left\{ \begin{array}{lcl}
                         (10^{0.7\log \lambda + 0.59})/0.6 &
                         \phantom{xxxxxx} & 0.01 <
                         \lambda \leq 2 \\
			 10.45 & \phantom{xxxxxx} &\lambda > 2
			  \end{array}
                  \right.
\end{equation}

\noindent and

\begin{equation}
\log \xi = \left\{ \begin{array}{lcl}
                  1.0 & \phantom{xxxxxx} & \log \lambda \leq -1 \\
		  2.0(\log \lambda +1) + 1 & \phantom{xxxxxx} & -1 < \log \lambda \leq
		  1.5 \\
		  6.0 & \phantom{xxxxxx} & \log \lambda > 1.5.
		  \end{array}
                  \right.
\end{equation}
The \fe\ equivalent width (EW) and line flux are measured for each $\lambda$ and then averaged over the
$\lambda$-distribution to obtain mean values for that particular X-ray
luminosity. The accretion disk is not expected to be optically thick
to X-rays when $\lambda \leq 0.01$ \citep{ny95}, and in those cases the \fe\
line EW and line flux are set to zero\footnote{Observations of Galactic black hole binaries seem to
  indicate that inner edge of the disk is a function of $\lambda$
  \citep[e.g.,][]{gdp08}, even when $\lambda > 0.01$, although other observations imply a more
  stable inner radius \citep[e.g.,][]{mill06}. If such an effect
  occurs in AGN disks, then the number density of relativistic lines
  from the lower $\lambda$ sources will be reduced. However, there is
  currently no observational evidence of such an effect in Seyfert galaxies or quasars.}. These calculations are repeated for 40 different
redshifts between $z=0.05$ and $z=2$, with the black hole mass
distribution moving to larger masses as $z^{0.3}$ \citep{lab09}.

It is currently unknown where the X-ray source is located in relation
to the accretion disk in AGNs, and so we consider two limiting cases for
$R$. The first is to assume $R=1$, a commonly
observed value in measurements of broad Fe lines \citep{nan07}, for all
calculations of the \fe\ EW and line flux. This is equivalent to
assuming the accretion disk subtends a solid angle of $\sim 2\pi$ as
viewed from the X-ray source. The illuminating power-law and reflection spectra are simply added
together to form the $R=1$ spectra. The second case assumes that $R$ is
inversely proportional to $\lambda$, which is a prediction of models
of relativistic light bending in AGNs where the X-ray source is
located on the axis of the central black hole \citep{mf04}. The relationship
between $R$ and $\lambda$ is determined from the \textit{Suzaku}
measurements of the Seyfert~1 galaxy \mcg\ that assumed a variable
$R$ \citep{min07}. A black hole mass of
$4\times 10^6$~M$_{\odot}$ \citep{mch05}, and a 2--10~\kev\ bolometric
correction of $50$ \citep{vf07} was used to calculate $\lambda$ for the
two states of \mcg. A straight-line fit yielded:
\begin{equation}
R = \left\{ \begin{array}{lcl}
                  8.0 & \phantom{xxxxxx} & \lambda \leq 0.09 \\
                  -8.5\lambda+8.8 & \phantom{xxxxxx} & 0.09 < \lambda \leq 0.92 \\
                  1.0 & \phantom{xxxxxx} & \lambda > 0.92.
                \end{array}
                \right.
\end{equation}
In both cases, the \fe\ equivalent widths and line fluxes are computed
by integrating over the total spectrum from $6$~\kev\ to
$7.1$~\kev. For the EW calculation, the continuum underneath the line
is estimated by fitting a straight line to the spectrum between those
two energies.  To calculate the line flux an emitting area from the
disk had to be assumed. Following the line emission model of
\citet{nan07}, we assumed that each line was emitted
from $1.235$--$6$~$r_g$ with a flat emissivity index, and from
$6$--$400$~$r_g$ with an emissivity index of $-3$. Here
$r_g=GM_{\mathrm{BH}}/c^2$ is the gravitational radius of a black hole
with mass $M_{\mathrm{BH}}$, $G$ is the gravitational constant and $c$
is the speed of light. 

\section{Results}
\label{sect:res}
 At any $z$, multiplying the observed X-ray luminosity function by the
  slope of the computed luminosity-EW relationship results in the \fe\
  EW analogy to the AGN luminosity function, which we call the \fe\ EW
  function. Figure~1 plots this function at four different redshifts.
There are strong peaks in the $R=1$ function at EWs of $\sim 110$~eV
  and $\sim 300$~eV at all redshifts that correspond to strong neutral
  and ionized \fe\ lines, respectively. In contrast, the strongest
  peak in the light-bending model is at an EW of $\sim 500$~eV which
  corresponds to neutral line emission that is enhanced by strong
  reflection. The $R=1$ model indicates that \fe\ lines with EWs $<
  110$~eV will dominate the space density at all redshifts, while the
  light-bending model predicts a significant space density of very
  strong \fe\ lines with EWs up to $\sim 500$~eV.

Each panel in Fig.~1 also gives the average \fe\ EW at the specified redshift
for the two different models. The average EW for the $R=1$ model is
$\sim 50$~eV for $z < 1$ and then rises to $\sim 70$~eV at higher
redshifts. In the extreme light-bending model, the average \fe\ EW is
$\sim 200$~eV over the entire redshift range. 

Figure 1 also shows a clear difference in the expected
density of relativistic \fe\ with EWs $> 100$~eV between the $R=1$ and
the light-bending models. Although there exists up to a factor of 10
scatter in the observed relationships connecting $\Gamma$, $\xi$,
$A_{\mathrm{Fe}}$ and $\lambda$ \citep{rye09,ith07,nt07}, this scatter
could in no way explain the three order of magnitude difference
between the $R=1$ and light-bending models. Moreover, the decrease in
high EW \fe\ lines in the $R=1$ model is driven by the observed X-ray
luminosity function.  A recent compilation by \citet{nan07} found that the majority of broad \fe\ components in
nearby Seyferts have EWs less than 100~eV, in agreement with the \fe\
function for $R=1$. At $z=0.05$, the light-bending model predicts a
space density of AGNs with \fe\ EWs $> 100$~eV that is 5$\times$
larger than the $R=1$ model (this ratio drops to $\approx 1$ at
$z=1$); for EWs $> 200$~eV the space-density is $235\times$ larger in
the light-bending model (this ratio drops to $\sim 150$ at
$z=1$). Observations indicate that such intense lines are
rare \citep{nan07}, despite the fact that they would be the easiest to
detect. However, our calculations assumed only one disk-like reflector
in an AGN, but this may be violated for AGNs in certain
states \citep{fb02} where multiple disk reflections can boost the
observed EW. Such a scenario seems to be required to explain the very
large \fe\ EW observed from 1H~0707-495 \citep{fab09}.

The predicted sky density distribution of relativistic \fe\ lines in
the range $z=0.05$--$0.5$ is shown in Figure~2. This plot dramatically illustrates that, in the $R=1$ scenario, AGNs
with \fe\ EWs $> 200$~eV will be very rarely observed between $z=0.05$
and $0.5$. On the other hand, if the X-ray source in AGNs is located
above the black hole, as in the extreme light-bending model, then such
large EWs would be common, with sky densities of $\sim
500$~deg$^{-2}$. If this model was generally true, then a near flat
distribution of EWs should be observed, while, in contrast, the
majority of detected broad \fe\ lines have EWs$ \leq
100$~eV \citep{nan07}, in agreement with the $R=1$ curve. The bottom
line is that, unless extreme light-bending is increasing the value of
$R$, relativistic \fe\ lines with EWs~$\ga 300$~eV can only be produce
by ionized accretion disks with large Fe abundances. Such a situation
is most common when an AGN has a large Eddington ratio, and such AGNs
are rare at $z < 1$.

One can use these results to plan future \fe\ line surveys performed
by the proposed \textit{International X-ray Observatory} (\textit{IXO}). Figure~3 plots the predicted sky density of
\fe\ lines greater than a certain line flux in AGNs between $z=0.05$
and $2$. The plot shows that both the $R=1$ scenario and the extreme
light-bending model predict that a small number of very bright lines
with fluxes $> 10^{-12}$~erg~cm$^{-2}$~s$^{-1}$. For reference, the
\fe\ line in the Seyfert 1 galaxy \mcg\ is one of the brightest known
and has a line flux of $\sim 2\times 10^{-12}$~erg~cm$^{-2}$~s$^{-1}$
\citep{ve01}. This plot predicts that there should only be $\sim 8$
AGNs over the entire sky with \fe\ lines that have fluxes equal to or
greater than \mcg, in agreement with the observed scarcity of such
lines. At lower fluxes, the predictions of the two models diverge with
the difference approaching 2 orders of magnitude at fluxes of $\sim
10^{-15}$~erg~cm$^{-2}$~s$^{-1}$. Such a difference will be easily
distinguished by future surveys by \textit{IXO}, but could be measured
by a careful analysis of the \textit{XMM-Newton} and \textit{Suzaku}
archives.

\section{Concluding Remarks}
\label{concl}
This work has shown, for the first time, the cosmic density and EW
distribution of relativistic \fe\ lines. Sensitive observations of
these line profiles are vital as they allow
measurements of fundamental parameters such as the spin of the central
black hole. Our results validate the previous and current
results from \textit{XMM-Newton} and \textit{Suzaku} that have shown
that the majority of relativistic \fe\ lines have EWs$< 100$~eV \citep{nan07}. Now that models of the intrinsic line EW and flux distributions are
available, detailed planning of future \fe\ surveys by \textit{IXO}
can be performed. The sensitivity of \textit{IXO} will push spin measurements beyond the
local universe and out to moderate redshifts, revolutionizing our
understanding of the cosmic evolution of black holes.

Finally, we have shown that the extreme light-bending model predicts
many more intense relativistic \fe\ lines than the $R=1$ model
(Figs. 1 and 2), in disagreement with current observational
constraints. This implies that, on average, the X-ray source in AGNs
does not lie on the axis of the black hole where light-bending would
be so extreme. It seems more likely that the X-ray source in average
AGNs lies above the accretion disk at a radius of several $r_g$ from
the black hole where light bending is less severe. Comparing data
from future surveys of \fe\ lines with plots such as Figure~3 should
allow a measurement of the average radial displacement of the X-ray
source. Of course, this conclusion does not preclude the possibility
that extreme light-bending occurs in a few rare sources.

\acknowledgments
The author thanks J.M. Miller and C.S. Reynolds for comments and
advice. This paper is dedicated to the memory of Don Ballantyne.

\clearpage

\begin{figure}[t!]
\begin{center}
\includegraphics[width=1.0\textwidth]{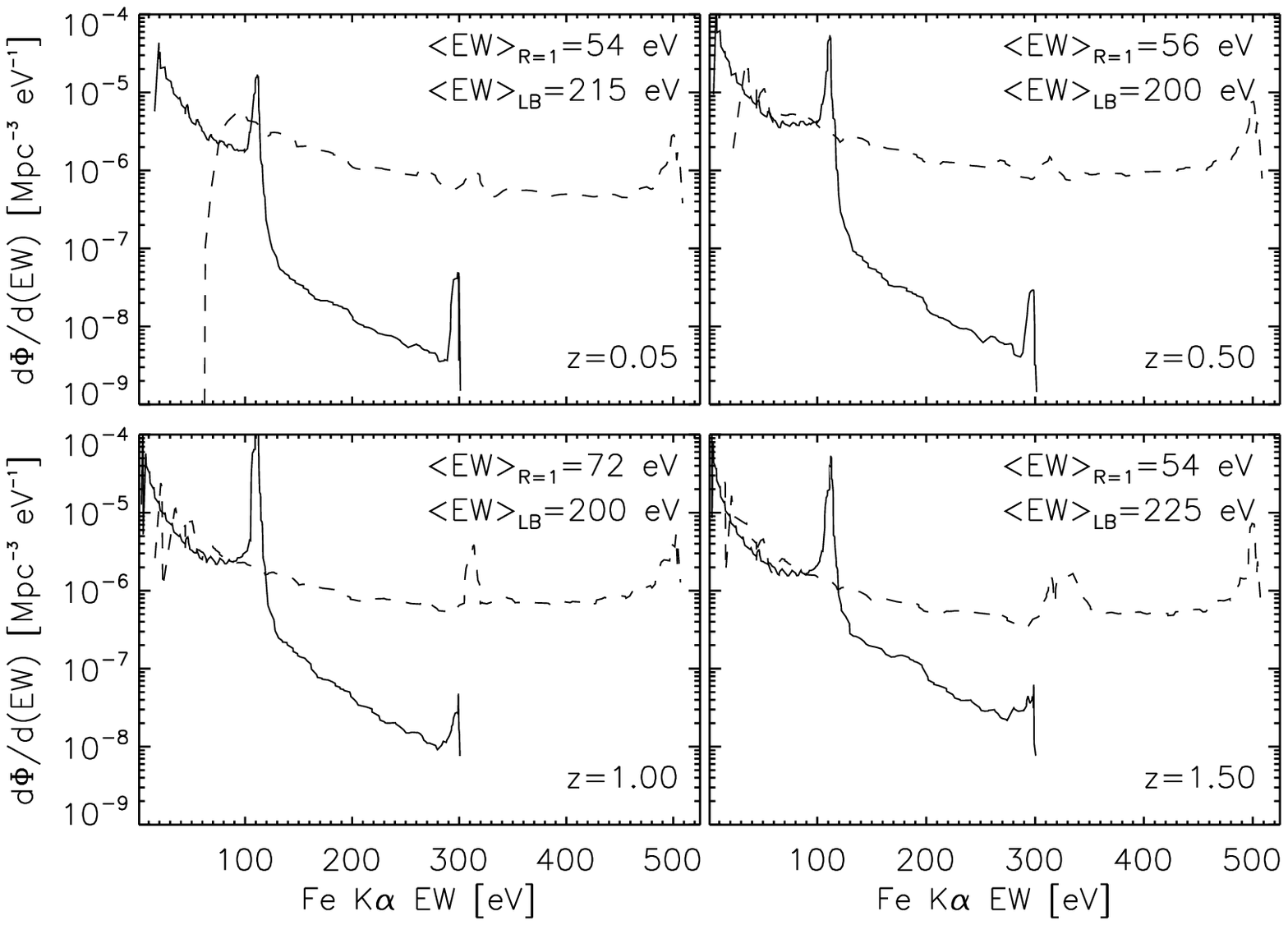}
\caption{The rest-frame \fe\ equivalent width function at $z=0.05$,
  $0.50$, $1.0$ and $1.5$. The solid line plots the function assuming
  a constant $R=1$, while the dashed line follows the strong
  light-bending prescription. The peaks in the curves indicate EWs
  that are most common at that redshift. For the $R=1$ model the peaks
  correspond to neutral \fe\ lines (at $\sim 110$~eV) and ionized
  reflection (at $\sim 300$~eV). In the light-bending model, the lines
  are always produced by neutral iron. The
  roll-over at low EWs in the light-bending model at $z=0.05$
  corresponds to ionized \fe\ lines from luminous quasars (which have
  very low space density at that redshift). The average EWs for both
  models are also indicated in the plots. The curves have been smoothed for clarity.}
\end{center}
\end{figure}

\clearpage

\begin{figure}
\begin{center}
\includegraphics[width=0.85\textwidth]{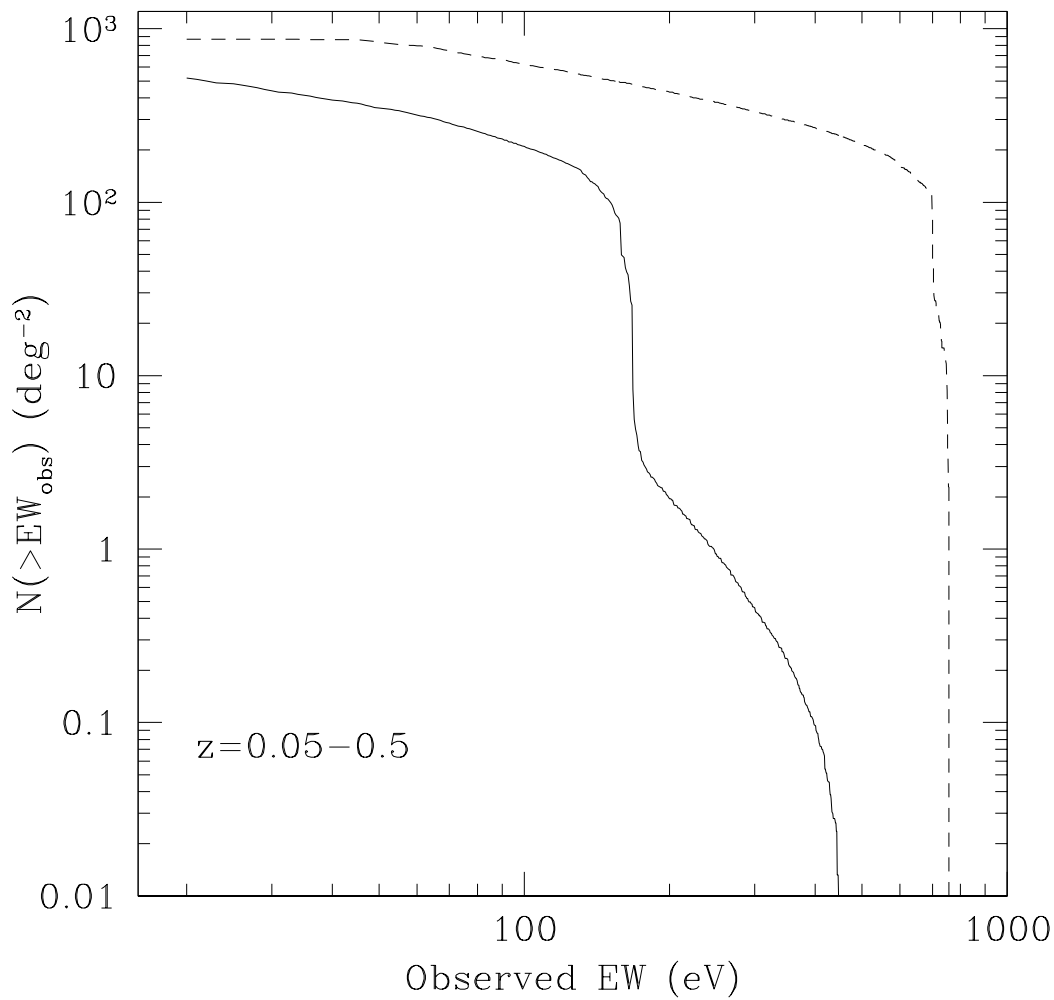}
\caption{The expected sky density of relativistic \fe\ lines from AGNs in the
  range $z=0.05$--$0.5$ with EWs greater than a certain observed
  value. The solid line plots the distribution from the models with a
  constant $R=1$, while the dashed line shows the predictions from the
  light-bending model where $R$ is inversely proportional to
  $\lambda$. The EWs in this plot are measured in the observed frame.}
\end{center}
\end{figure}

\clearpage

\begin{figure}
\begin{center}
\includegraphics[width=0.85\textwidth]{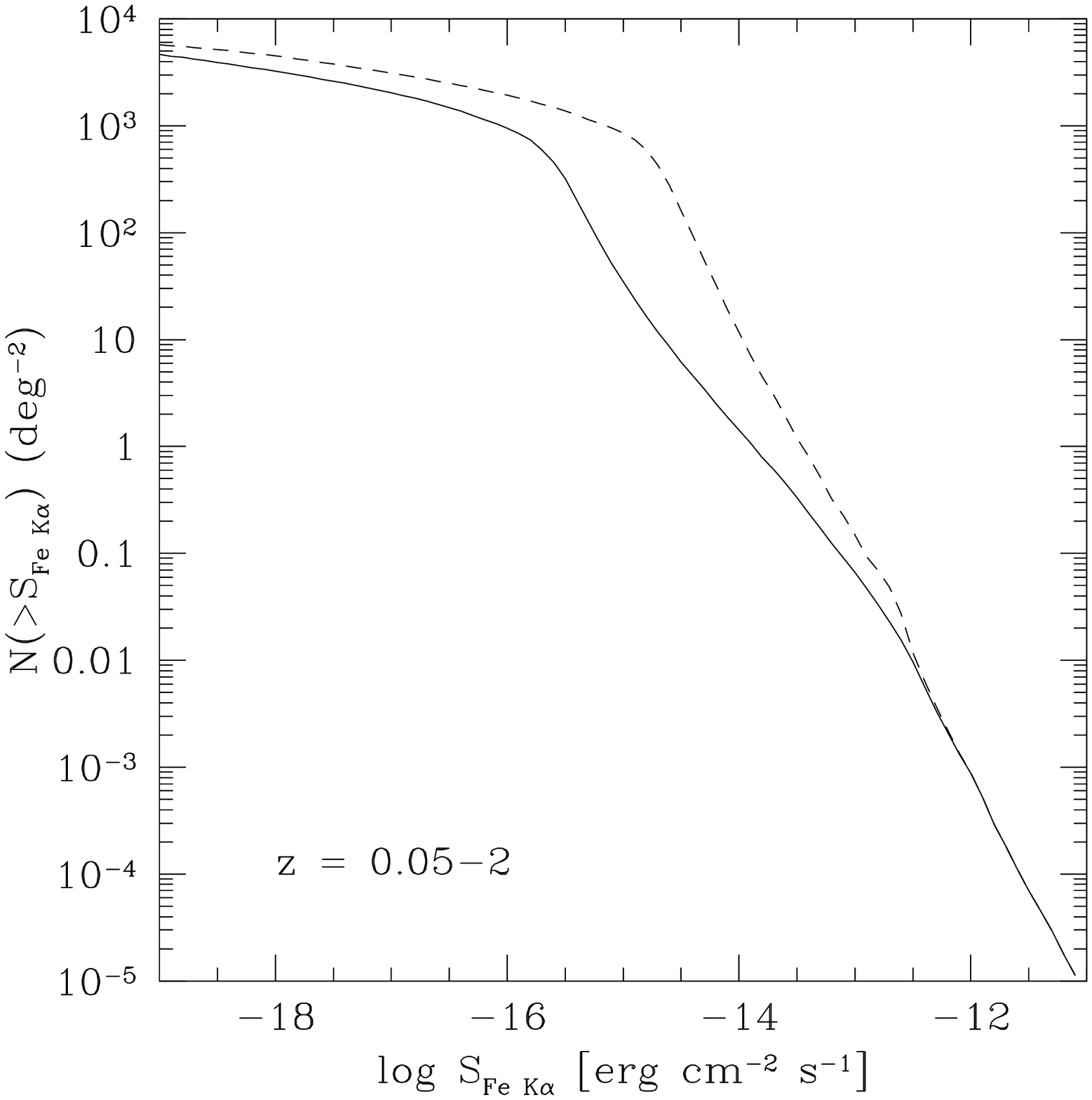}
\caption{The expected sky density of relativistic \fe\ lines from AGNs
  in the range $z=0.05$--$2$ with line fluxes greater than a certain
  value. The solid line plots the distribution from the models with a
  constant $R=1$, while the dashed line shows the predictions from the
  light-bending model where $R$ is inversely proportional to
  $\lambda$.}
\end{center}
\end{figure}

\end{document}